\documentclass[aps,prb,twocolumn,floatfix]{revtex4-1}

\usepackage[utf8]{inputenc}
\usepackage{amsfonts}
\usepackage{amsmath}
\usepackage{amssymb}
\usepackage{bbm}
\usepackage{amsthm}
\usepackage{empheq}
\usepackage{enumitem}
\usepackage{verbatim}
\usepackage[dvips]{graphicx}
\usepackage[dvips]{color}
\usepackage{subfigure}

\usepackage{hyperref}

\begin{document}

\title{Finite momentum condensate brought on by Zeeman field}
\author{Stephen Keeling$^{1}$ and Predrag Nikoli\'c$^{1,2}$}
\affiliation{$^1$Department of Physics and Astronomy, George Mason University, Fairfax, VA 22030, USA}
\affiliation{$^2$Institute for Quantum Matter at Johns Hopkins University, Baltimore, MD 21218, USA}
\date{\today}

\begin{abstract}
We study superfluid states in a two-dimensional fermionic attractive Hubbard model with Zeeman coupling to an external field. Focusing our attention on singlet pairing in both weak and strong coupling regimes, we reveal a rich phase diagram of finite momentum condensates which exhibits both Fulde-Ferrell and Larkin-Ovchinnikov orders at zero temperature. The latter are commensurate stripe states that spontaneously break a lattice symmetry; many stable ordering wavevectors are found as a function of particle density and Zeeman field. Stronger coupling significantly enhances the stability of the finite momentum condensates, but our numerical mean-field calculations underestimate the effect of fluctuations and indicate a possible localization near half-filling.
\end{abstract}

\maketitle

\section{Introduction}\label{intro}

It has been over fifty years since Fulde and Ferrell\cite{fulde1964superconductivity} (FF), and Larkin and Ovchinnikov \cite{larkin1965inhomogeneous} (LO) separately predicted the existence of spin-polarized superfluids in high magnetic fields. The FFLO superfluids feature Cooper pair condensation at finite wavevectors which allows a coexistence of spin-polarized unpaired fermions, for example at locations where the superfluid density is depleted. Cooper pairing at finite momenta has been envisioned in a wide variety of physical systems: dense quark matter within neutron stars \cite{alford2001crystalline}, atomic nuclei \cite{casalbuoni2004inhomogeneous}, heavy-fermion superconductors \cite{matsuda2007fulde}, and ultra-cold atomic gases \cite{revelle20161d, mitra2016phase}. However, the unambiguous detection of the FFLO phase has proven exceedingly difficult. Studies of heavy-fermion systems \cite{bianchi2003possible, radovan2003magnetic, bianchi2008superconducting, koutroulakis2010field} and layered organic materials \cite{beyer2012angle, lortz2007calorimetric, bergk2011magnetic, wright2011zeeman, mayaffre2014evidence} have in some cases shown promising indications of FFLO states. An FFLO state was reported in a trapped continuum 1D atomic gas \cite{liao2010spin} where its defining long-range order cannot survive the fluctuations even at zero temperature. While trapped ultra-cold gases of atoms are a very promising system for discovering FFLO states, their observation in higher dimensions has been plagued by phase separation \cite{zwierlein2006fermionic, partridge2006pairing}.

The rapid experimental progress in the field of cold atoms has motivated diverse theoretical explorations of FFLO states \cite{Kinnunen2018}. Exact analytical and numerical studies have been conducted in one-dimensional systems \cite{buzdin1983phase, machida1984superconductivity, buzdin1987nonuniform, orso2007attractive, yang2001inhomogeneous, feiguin2007pairing, batrouni2008exact}, while the mean-field approximation is the most widely used method in higher dimensions. The mean-field phase diagram of continuum systems contains only a thin sliver of FFLO states in three dimensions \cite{sheehy2006bec, Yoshida2007}, and the FFLO region is noticeable and broad in two dimensions \cite{shimahara1998structure, sheehy2015fulde}. Fluctuations typically challenge the stability of simple mean-field FFLO states in continuum systems \cite{Shimahara1998b, Radzihovsky2009, parish2013highly, yin2014fulde}, but may spare more complicated multi-wavevector FFLO states \cite{shimahara1998structure, Shimahara1998b}, or evolve them into less conventional states, e.g. nematic \cite{Radzihovsky2009}. On the other hand, lattice systems provide a new mechanism for stabilizing the FFLO superfluids both in two dimensions \cite{Koponen2008, Cui2009, wolak2012pairing, Chiesa2013, Troyer2016, baarsma2016larkin} and three dimensions \cite{Koponen2007, Koponen2008, loh2010detecting}.

In this paper, we analyze singlet pairing in the attractive Hubbard model on the square lattice subjected to an external Zeeman field. We calculate the mean-field phase diagram numerically and reveal a rich structure of finite momentum condensates (FMC) in the intermediate and strong coupling regimes, both at zero and finite temperatures. In contrast to previous mean-field studies of this model \cite{Koponen2008, Chiesa2013, baarsma2016larkin}, we demonstrate that commensurate pair density wave orders (LO states) can form with many different ordering wavevectors, aligned either with a lattice direction or a lattice diagonal. Such states are embedded inside the dome of finite momentum condensates in the phase diagram, bordering the uniform superfluid state (USF). An intermediate plane wave (FF) phase is typically found at highest Zeeman fields, bordering the normal state -- it becomes incommensurate at finite temperatures. Strong pairing interactions are seen to enhance the variety and stability of FMC states at the mean-field level, but the effects of fluctuations are also expected to be more severe. The elaborate scrutiny of fluctuations in this model \cite{wolak2012pairing, Troyer2016} has confirmed the stability of FFLO phases without being able yet to address the specific ordering patterns deep in the superfluid state. Hence, our results shed some new insight into this problem and invite further study. We anticipate that the discrete symmetries of the lattice will protect broken translational and rotational symmetries of the FMC states even when the U(1) symmetry is restored (e.g. at finite temperatures). We speculate that the most dramatic effect of fluctuations can be the localization of Cooper pairs in the strong coupling regime, leading to unconventional spin-polarized Mott insulators. Our discovery of many different commensurate pair density waves is related (via a particle-hole transformation \cite{Kinnunen2018}) to the predicted existence of unconventional Mott insulators in the repulsive Hubbard model relevant to cuprate high-$T_{\textrm{C}}$ superconductors \cite{balents05, balents05a}.

The rest of the paper is organized as follows. Section \ref{secModel} introduces the model we study, and describes the representation of the Bogoliubov-de Gennes Hamiltonian that allows us to analyze a variety of order parameters. Section \ref{secPD} explains the numerical procedure and presents the mean-field phase diagram of FMC states at zero and finite temperatures. We qualitatively discuss the fluctuation effects in Section \ref{secFluct}, and summarize the conclusions in Section \ref{secConclusions}. Some technical details about the numerical procedure are given in the Appendix.

\begin{figure*}[t!]
\includegraphics[height=1.7in]{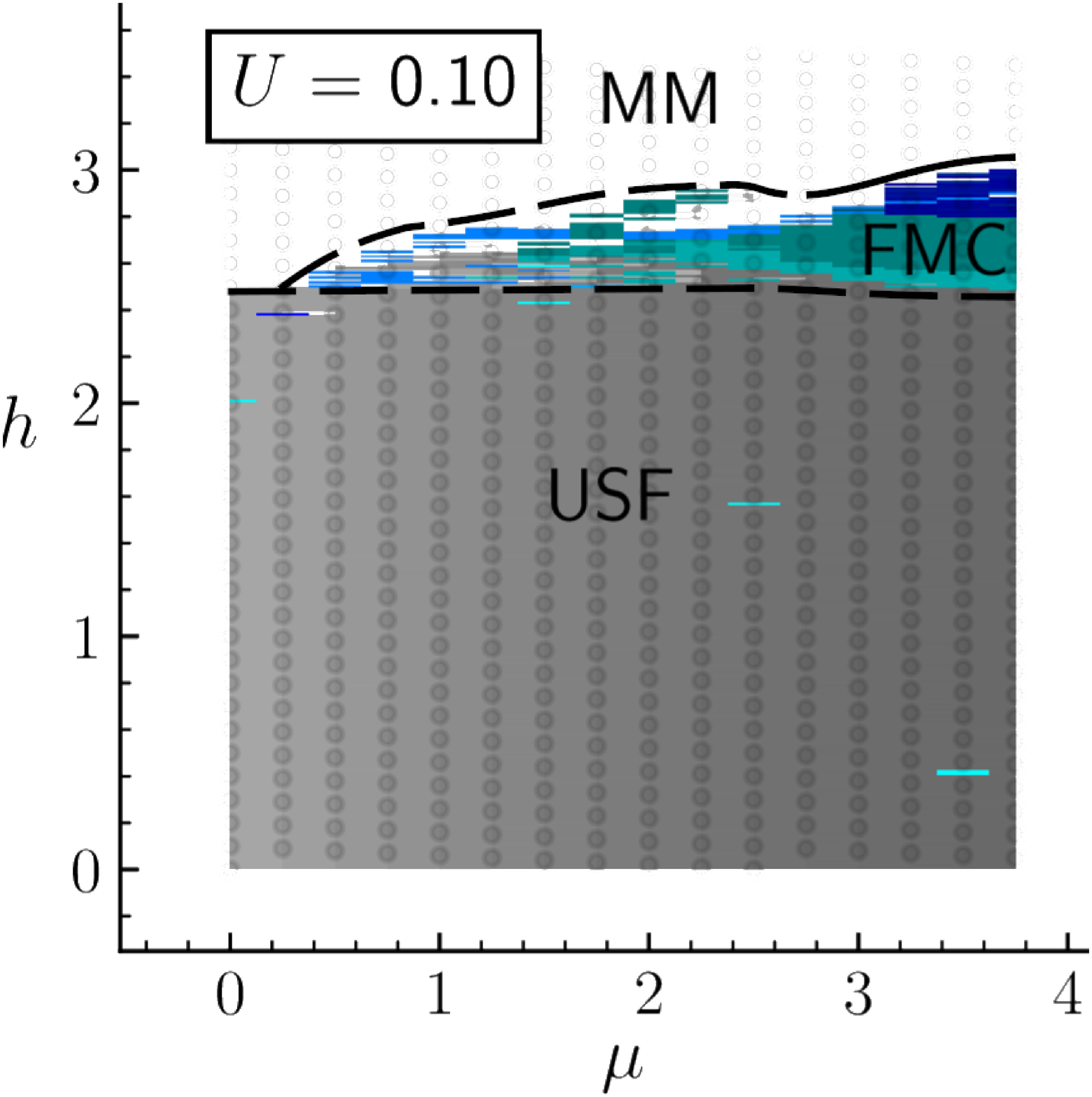} \hskip -0.05in
\includegraphics[height=1.7in]{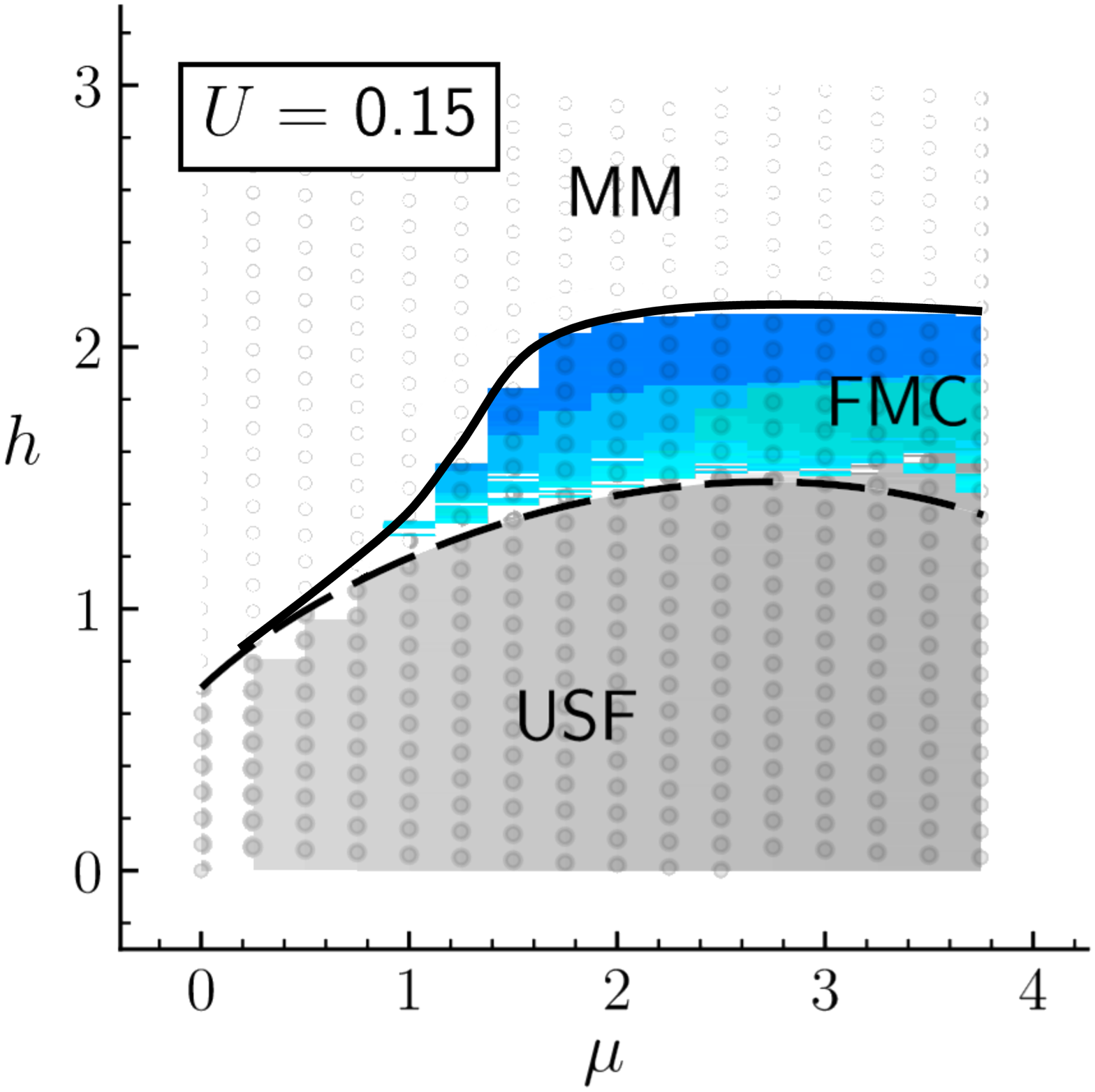} \hskip -0.05in
\includegraphics[height=1.7in]{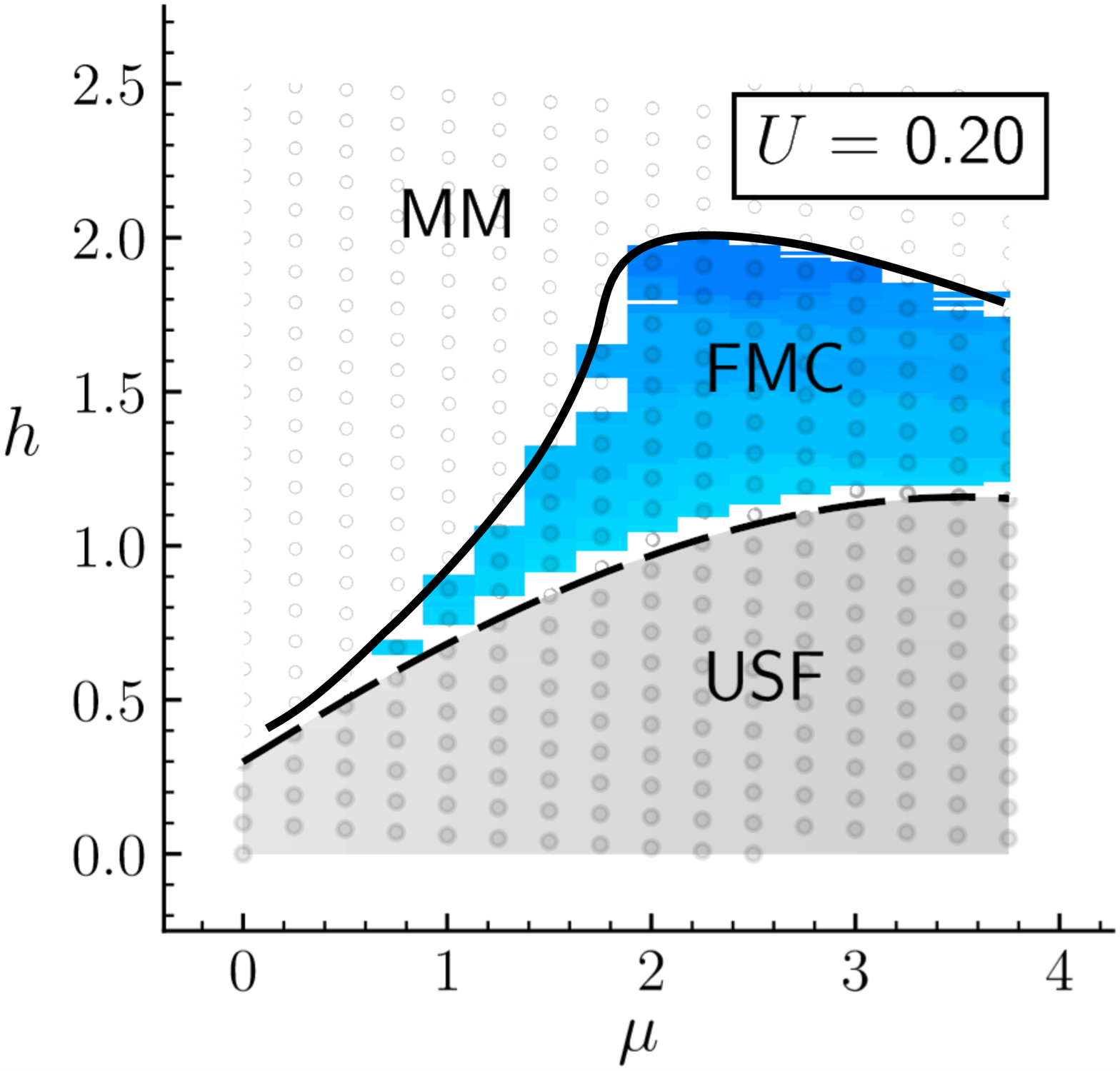} \hskip -0.05in
\includegraphics[height=1.7in]{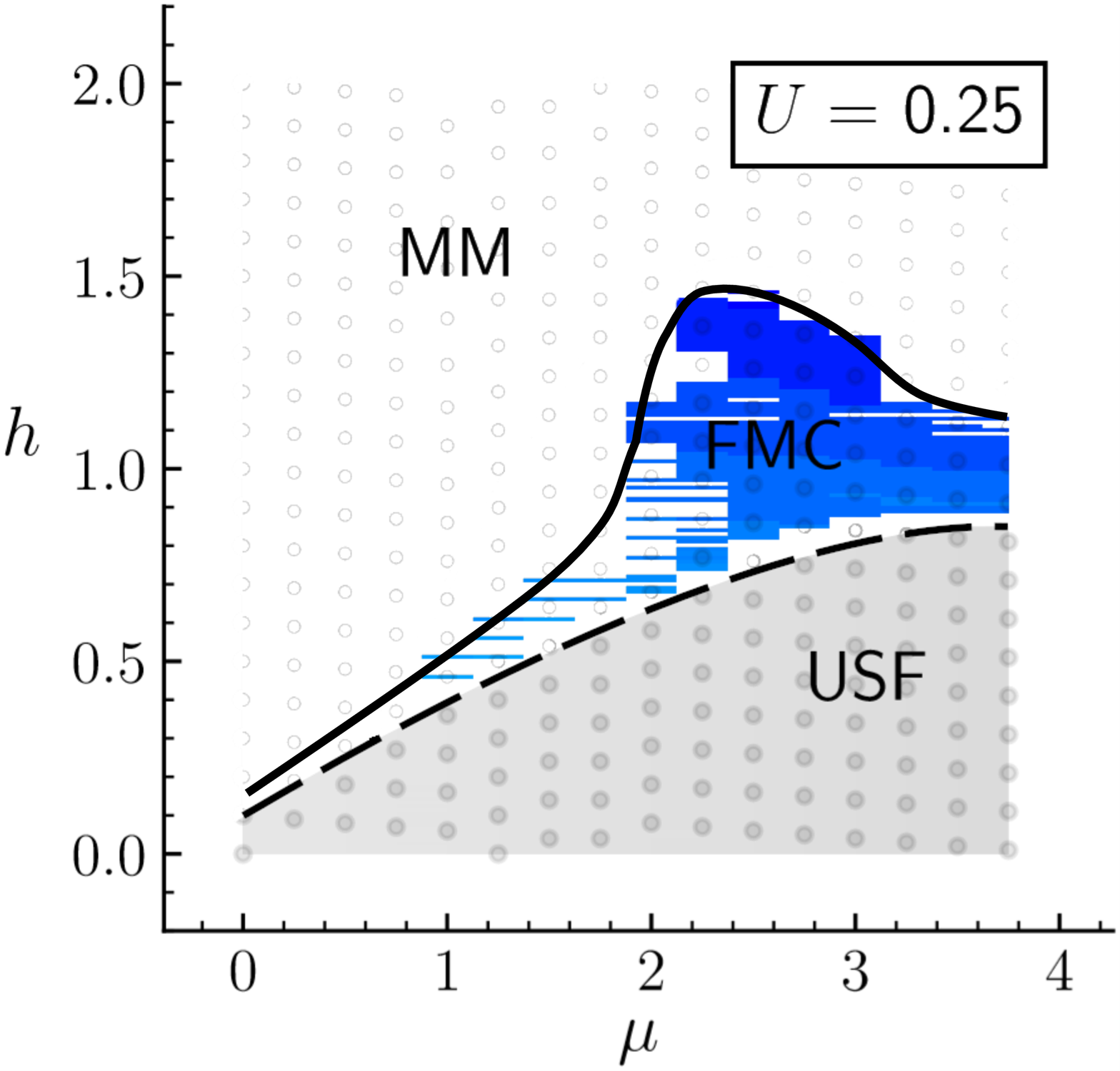}
\caption{\label{fig:phdiag:mu-h}The overall mean-field phase diagram of the model (\ref{modelHamiltonian}), expressed in units $t=a=1$. Uniform superfluid (USF) is shaded gray, finite-momentum condensates (FMC) are shaded blue, and the normal magnetized metallic (MM) state is unshaded. The shading gradient represents the scaled superfluid density (order parameter magnitude), and the color of the FMC regions is determined by mixing the blue pigment for $Q_x$ and green pigment for $Q_y$ component of the ordering wavevector ${\bf Q}=(Q_x,Q_y)$. Dashed thick lines representing first-order phase transitions and solid lines representing second-order transitions are hand-drawn to guide the eye (not all transitions are indicated, see Fig.\ref{fig:phdiag:mu-h:fmc}). Circles mark every tenth sample point in numerical calculations.}
\end{figure*}

\section{Competing superconducting orders}

\subsection{Model}\label{secModel}

We model a gas of ultra cold fermions on a two-dimensional square lattice with the following Hamiltonian:
\begin{align} \label{modelHamiltonian}
    H = &\sum_{\mathbf{k}\sigma}{ \xi_{\mathbf{k}}^{\phantom{\dagger}}c_{\mathbf{k}\sigma}^{\dagger}c_{\mathbf{k}\sigma}^{\phantom{\dagger}}}
      - h \sum_{  \mathbf{k}}{ \left( c_{\mathbf{k}\uparrow}^{\dagger}c_{\mathbf{k}\uparrow}^{\phantom{\dagger}} -
         c_{\mathbf{k}\downarrow}^{\dagger}c_{\mathbf{k}\downarrow}^{\phantom{\dagger}} \right)} \\
	      &+ \sum_{\mathbf{q}} \sum_{  \mathbf{k}}{ \left( \Delta_{\mathbf{q}}^{\phantom{\dagger}} c_{\mathbf{k} \uparrow}^{\dagger}
	         c_{\mathbf{q}-\mathbf{k} \downarrow}^{\dagger} + h.c. \right) }
                         + U\sum_{\mathbf{q}}\Delta_\mathbf{q}^*\Delta_\mathbf{q}^{\phantom{x}} \ . \nonumber
\end{align}
All sums are restricted to wavevectors in the first Brillouin zone $|k_x|,|k_y|<\pi/a$ appropriate for the square lattice with lattice constant $a$. We define $\xi_{\mathbf{k}} = \epsilon_{\mathbf{k}} - \mu$ with the square lattice tight-binding dispersion
\begin{equation}\label{Dispersion}
\epsilon_{\mathbf{k}} = 2t \left( 2 - \cos k_x a - \cos k_y a\right) \ ,
\end{equation}
and work in the grand canonical ensemble with chemical potential $\mu$ and Zeeman field $h$. $U^{-1}$ parameterizes the on-site attractive interaction strength between fermions upon integrating out $\Delta_{\bf q}$ as a Hubbard-Stratonovich field in the appropriate path integral. The operators $c_{\mathbf{k} \sigma}^{\dagger}, c_{\mathbf{k} \sigma}^{\phantom{\dagger}}$ create and annihilate respectively a fermion with (pseudo)spin $\sigma \in {\uparrow, \downarrow}$ and crystal wavevector $\mathbf{k}$. The superfluid order parameter can have non-zero amplitudes $\Delta_{\mathbf{q}}$ at multiple wavevectors ${\bf q}$.

Simple periodic condensates in real space are characterized by a single commensurate ordering wavevector ${\bf Q}$:
\begin{equation}\label{OP}
    \Delta({\bf r}) = \sum_{n=0}^{q-1} \Delta_n e^{i n {\bf Qr}} \ ,
\end{equation}
where ${\bf r}$ is a position on the lattice and $q$ is the smallest integer that makes $q{\bf Q}$ a reciprocal lattice vector. The set of non-zero complex pairing amplitudes $\Delta_n$ determines the nature of the condensate. If $\Delta_n\neq 0$ for only one $0\le n<q$, the condensate is either a plane wave (Fulde-Ferrell) state ($n\neq 0$) or a conventional ``uniform'' state ($n=0$). In both cases, the state possesses translational symmetry (up to a global gauge transformation), but the plane wave states normally break the time-reversal (TR) symmetry. Multiple non-zero amplitudes $\Delta_n\neq 0$ produce pair density waves (Larkin-Ovchinnikov states) in real space, possibly with additional Cooper pair currents that violate time reversal. Larger ordering wavevectors $|{\bf Q}|$ correspond to smaller unit-cells of the periodic density waves. We generally find that periodic states with relatively small unit-cells are more stable, i.e. have a lower energy, than the states with arbitrarily larger unit-cells. This indicates that incommensurate density wave states are not equilibrium states in this model, as the size of their effective unit-cell diverges. In contrast, the translation-invariant plane wave states are never frustrated by the presence of a lattice, and can be incommensurate without a significant energy penalty. Therefore, we allow ${\bf Q}$ to be an arbitrary wavevector in the first Brillouin zone in the case of plane wave condensates.

\raggedbottom

Given a set of Hamiltonian parameters $\mu,h,U$, we determine the ground state by minimizing the free energy of the particles under variations of the ordering wavevector ${\bf Q}$ and condensate amplitudes $\Delta_n$. At zero temperature the free energy reduces to the ground state energy of this many-body system. In the mean-field approximation, the elementary excitations are Bogoliubov quasiparticles whose energy levels are populated up to the chemical potential in the ground state. Their energy spectrum is determined by rewriting (\ref{modelHamiltonian}) as
\begin{equation} \label{modelHamiltonian2}
H = \sum_{\bf k} \Psi_{\bf k}^{\dagger} H_{\textrm{BdG}}^{\phantom{x}}({\bf k}) \Psi_{\bf k}^{\phantom{\dagger}}
    + U\sum_{\mathbf{q}}\Delta_\mathbf{q}^*\Delta_\mathbf{q}^{\phantom{x}}
\end{equation}
and diagonalizing the Bogoliubov-de Gennes (BdG) Hamiltonian. Plane wave condensates $\Delta_n=\Delta_1 \delta_{n,1}$ admit a simple Nambu representation:
\begin{equation}\label{BdG-PWS}
\Psi_{\bf k} = \left( \begin{array}{c} c_{{\bf k}\uparrow} \\ c_{{\bf Q}-{\bf k}\downarrow}^{\dagger} \end{array} \right)
  \quad , \quad
H_{\textrm{BdG}}^{\textrm{(PWS)}} = \left( \begin{array}{cc}
  \xi_{\bf k} & \Delta_{1} \\
  \Delta_{1}^* & -\xi_{{\bf Q}-{\bf k}}
\end{array} \right) - h\times\mathbbm{1}
\end{equation}
for an arbitrary condensation wavevector ${\bf Q}$. Otherwise, we must enlarge the Nambu representation $q$ times and reduce the first Brillouin zone $q$ times for the sum over ${\bf k}$ in (\ref{modelHamiltonian2}) in order to capture arbitrary commensurate condensates of the form (\ref{OP}). If at each wavevector ${\bf k}$ of the reduced zone $\textrm{1BZ'}$ we represent the enlarged Nambu spinor as
\begin{widetext}
\begin{eqnarray}
\Psi_{\bf k}^T &=& \left( \begin{array}{ccccccccc}
  c_{{\bf k}\uparrow}^{\phantom{\dagger}} & c_{-{\bf k}\downarrow}^{\dagger} & c_{{\bf Q}+{\bf k}\uparrow}^{\phantom{\dagger}} & c_{-{\bf Q}-{\bf k}\downarrow}^{\dagger} &
  c_{2{\bf Q}+{\bf k}\uparrow}^{\phantom{\dagger}} & c_{-2{\bf Q}-{\bf k}\downarrow}^{\dagger} & \cdots &
  c_{(q-1){\bf Q}+{\bf k}\uparrow}^{\phantom{\dagger}} & c_{-(q-1){\bf Q}-{\bf k}\downarrow}^{\dagger}
\end{array} \right)
\nonumber \\[0.1in]
\end{eqnarray}
then the BdG Hamiltonian is:
\begin{eqnarray}\label{BdG-PDW}
H_{\textrm{BdG}}^{\textrm{(PDW)}} &=& \left( \begin{array}{ccccccccc}
  \xi_{\bf k} & \Delta_0   & 0 & \Delta_{q-1}   & 0 & \Delta_{q-2}   & \cdots   & 0 & \Delta_1  \\
  \Delta_0^* & -\xi_{-{\bf k}}  & \Delta_1^* & 0   & \Delta_2^* & 0   & \cdots   & \Delta_{q-1}^* & 0  \\
  0 & \Delta_1   & \xi_{{\bf Q}+{\bf k}} & \Delta_0   & 0 & \Delta_{q-1}    & \cdots   & 0 & \Delta_2  \\
  \Delta_{q-1}^* & 0   & \Delta_0^* & -\xi_{(q-1){\bf Q}-{\bf k}}  & \Delta_1^* & 0    & \cdots   & \Delta_{q-2}^* & 0  \\
  0 & \Delta_2   & 0 & \Delta_1   & \xi_{2{\bf Q}+{\bf k}} & \Delta_0   & \cdots   & 0 & \Delta_3  \\
  \Delta_{q-2}^* & 0   & \Delta_{q-1}^* & 0   & \Delta_0^* & -\xi_{(q-2){\bf Q}-{\bf k}}  & \cdots   & \Delta_{q-3}^* & 0  \\
  \vdots & \vdots & \vdots & \vdots & \vdots & \vdots   & \ddots   & \vdots & \vdots \\
  0 & \Delta_{q-1}   & 0 & \Delta_{q-2}   & 0 & \Delta_{q-3}   & \cdots & \xi_{(q-1){\bf Q}+{\bf k}} & \Delta_0 \\
  \Delta_1^* & 0   & \Delta_2^* & 0   & \Delta_3^* & 0   & \cdots & \Delta_0^* & -\xi_{{\bf Q}-{\bf k}}
\end{array} \right) - h\times\mathbbm{1} \ .
\end{eqnarray}
\end{widetext}

Here we are using the fact that $q{\bf Q}$ is equivalent to ${\bf Q}$, and $\xi_{\bf k} = \xi_{-{\bf k}}$.

\subsection{Mean-field phase diagram}\label{secPD}

Exploiting the symmetry of the square lattice, we search for order parameters in the following categories:
\begin{enumerate}
    \item commensurate ``diagonal'' pair-density wave (PDW-D) with ordering wavevector $\mathbf{Q} = 2\pi(f, f)$;
    \item commensurate ``columnar'' pair-density wave (PDW-X) with ordering wavevector $\mathbf{Q} = 2\pi(f,0)$;
    \item incommensurate plane wave state (PWS) with an arbitrary condensation wavevector $\mathbf{Q} = (q_x,q_y)$.
\end{enumerate}
The energy spectrum $E_{n,{\bf k}}$ of Bogoliubov quasiparticles is obtained by diagonalizing (\ref{BdG-PDW}) in categories 1 and 2, or (\ref{BdG-PWS}) in category 3; the quantum number $n$ combines the spin and band of quasiparticles in the periodic potential induced by the PDW order parameter, and ${\bf k}$ lives in the reduced first Brillouin zone associated with the broken lattice symmetry. The free energy at temperature $T$
\begin{equation}\label{FE}
F = U\sum_{\mathbf{q}}\Delta_\mathbf{q}^*\Delta_\mathbf{q}^{\phantom{x}} - \sum_n\int\limits_{\textrm{1BZ}'}\frac{d^2 k}{(2\pi)^2}T\ln\left(1+e^{-E_{n,{\bf k}}/T}\right)
\end{equation}
reduces to the ground state energy at $T=0$. The order parameter is determined by minimizing this free energy. The search for PDW-D/X can be computationally expensive, so we have taken advantage of the particle-hole symmetry of the model and focused on the $\mu<4t$ region.

The phase diagram of superfluid states, obtained by comparing the lowest energy states of category 1,2,3 calculations, is shown in Fig.\ref{fig:phdiag:mu-h}. The first striking feature is the large parameter space of finite momentum condensates (FMC) consisting of both PDW and PWS. The lattice system has stabilized PDW states for sufficiently large particle densities and Zeeman fields, in a broad range of interaction couplings $U$. The boundary between FMC and normal phases features a characteristic bulge due to Van Hove singularities \cite{Koponen2008} in the weak coupling regime $Ut\ge 0.2$, but this feature becomes washed out in the strong coupling limit $Ut\le0.15$.

\begin{figure*}
\includegraphics[height=2.5in]{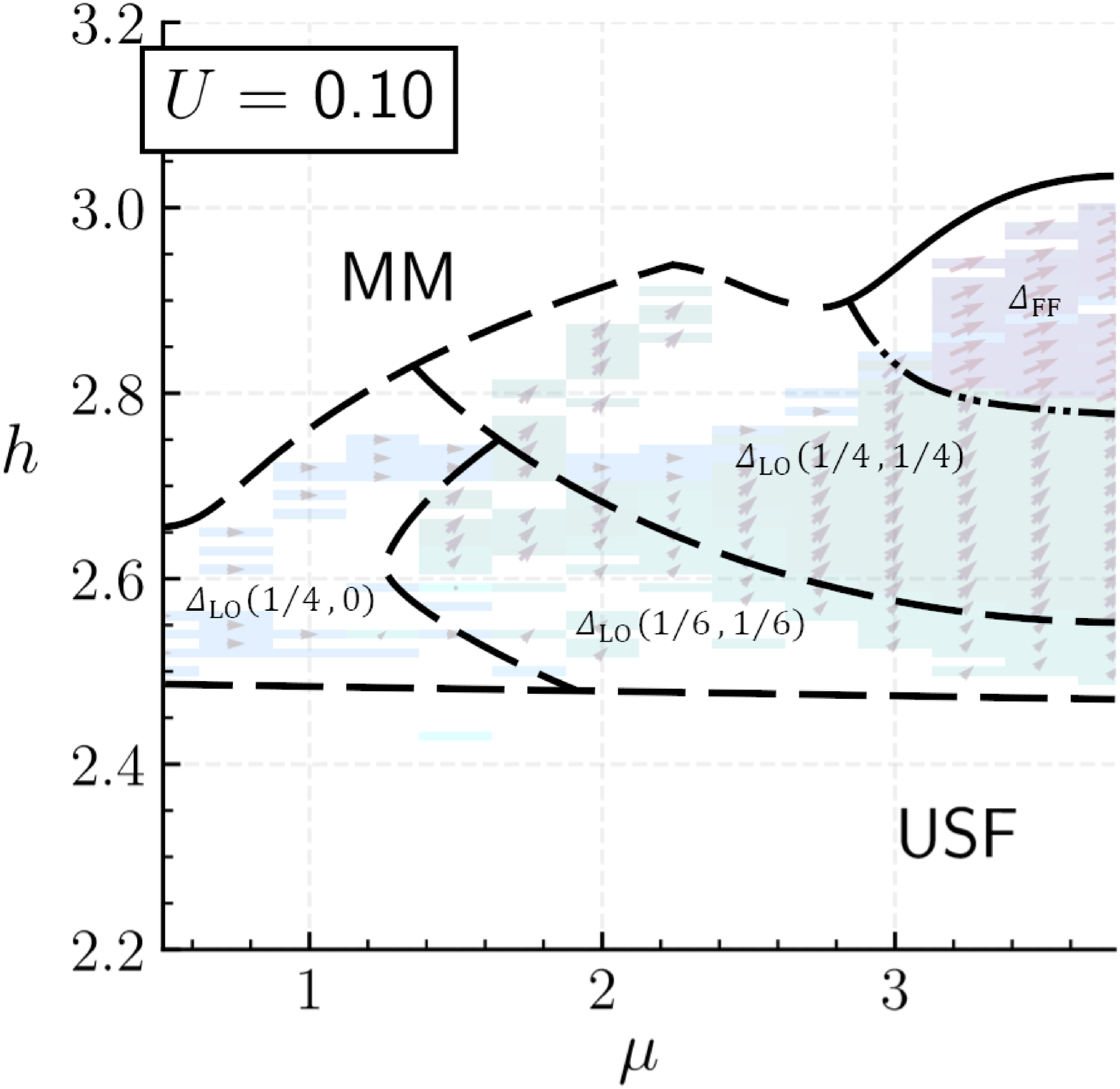} \hskip 0.5in
\includegraphics[height=2.5in]{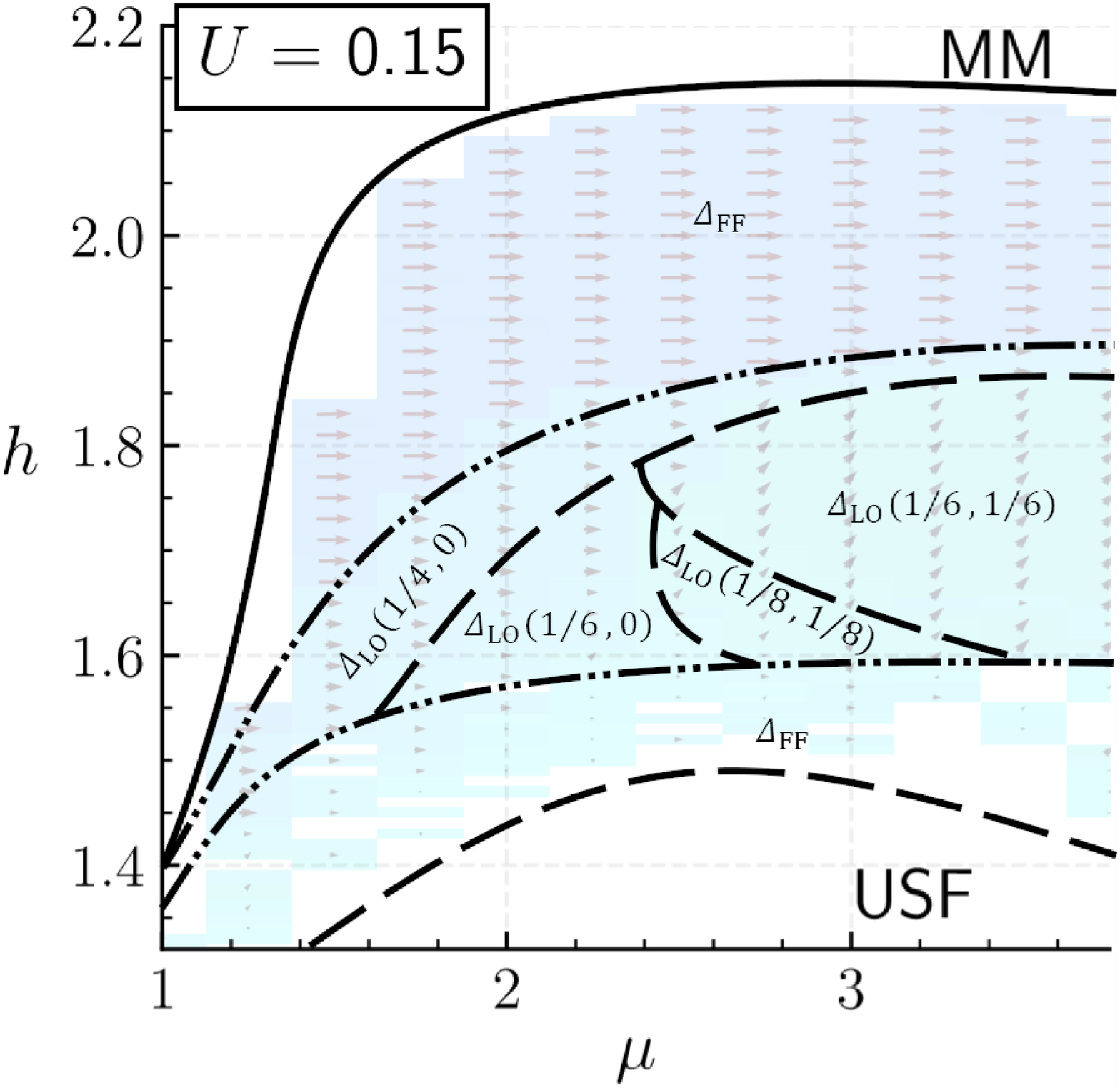}
\includegraphics[height=2.5in]{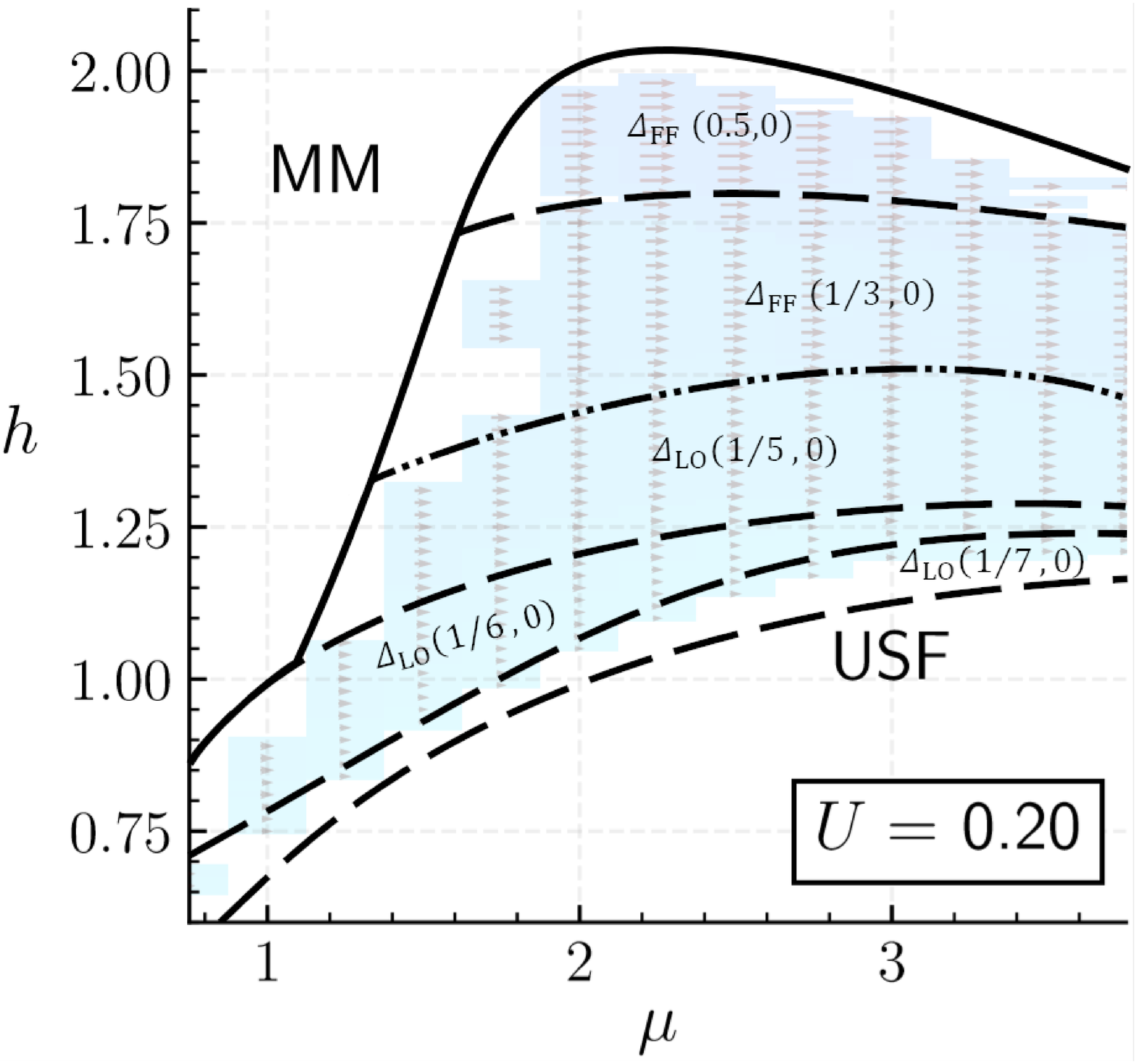} \hskip 0.5in
\includegraphics[height=2.5in]{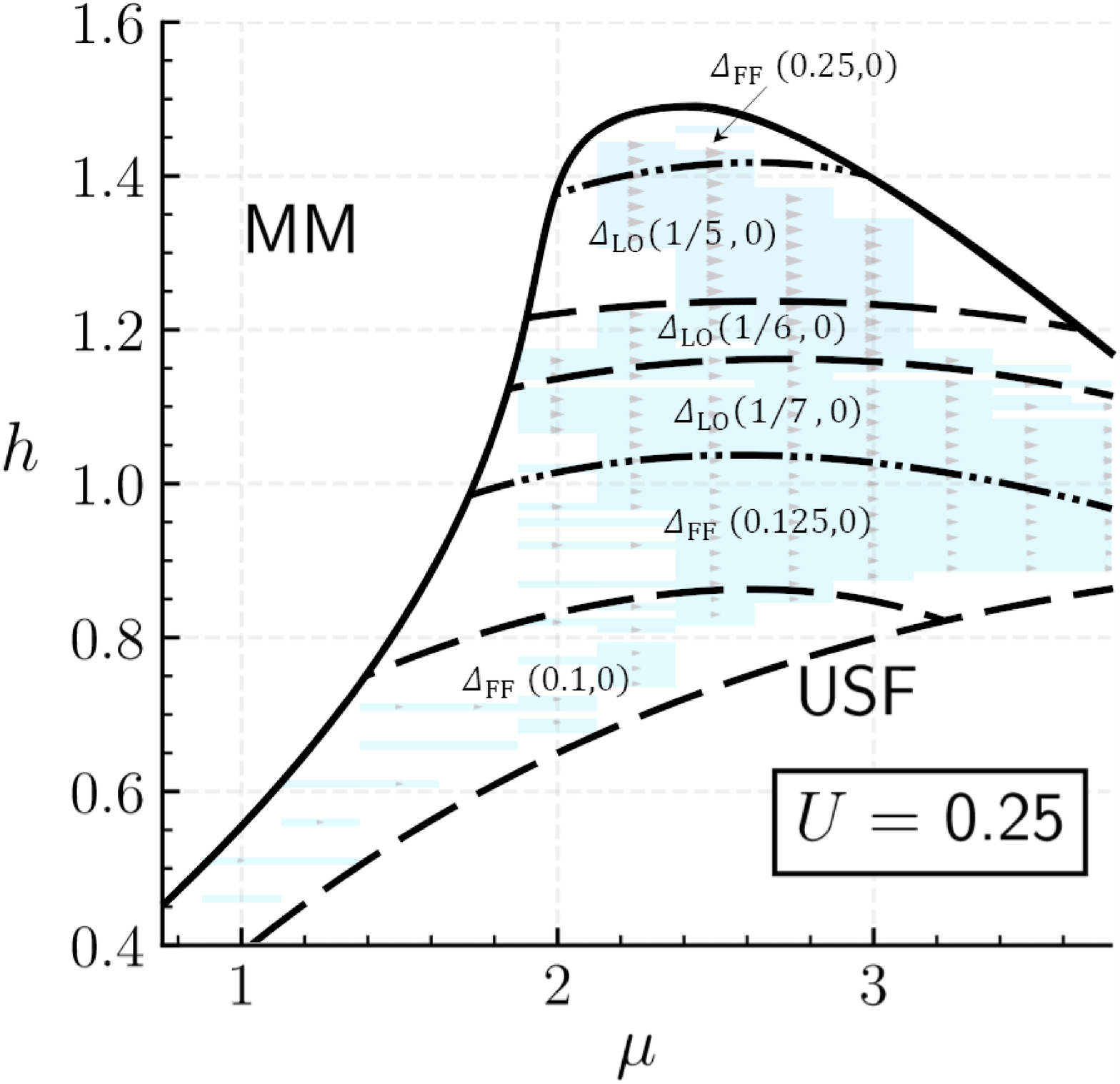}
\caption{\label{fig:phdiag:mu-h:fmc}Detailed phase diagram of the FMC regions featured in Fig.\ref{fig:phdiag:mu-h}. FMC phases are labeled by $\Delta_{\textrm{FF}}$ for plane-wave (PWS, Fulde-Ferrell) orders and $\Delta_{\textrm{LO}}$ for pair-density wave (PDW, Larkin-Ovchinnikov) orders, followed by the ordering wavevector ${\bf Q}=2\pi(q_x,q_y)$ in parenthesis. If an ordering wavevector component $q_x, q_y$ was found to be commensurate with the lattice (always the case in LO phases), it is denoted with an exact fraction. Some FF phases, with an omitted wavevector label, feature an ordering wavevector whose components smoothly evolve and grow with increasing $\mu$ or $h$. Arrows in the background also indicate the orientation and magnitude of ${\bf Q}$. Different FMC are separated by first-order transitions indicated with dashed lines, while the final transition to the normal state is typically second-order and indicated with a solid line, except in the strongest coupling regime. Other conventions are the same as in Fig.\ref{fig:phdiag:mu-h}.}
\end{figure*}

A closer look at the region of FMC, shown in Fig.\ref{fig:phdiag:mu-h:fmc}, reveals a phase space rich in competing orders. The FMC appears at particle densities away from half-filling for weak Zeeman fields. The FMC region broadens with increasing particle density, but reaches a peak\cite{baarsma2016larkin} for $Ut \sim 0.2-0.25$. The ordering wavevector magnitude $|{\bf Q}|$ generally grows with the Zeeman field as expected of FFLO states. The phase transitions between different condensed states are generally found to be first order, including the transition between USF and FMC. The transition between the PWS and the normal magnetized metal (MM) state is typically second order as in the continuum limit \cite{sheehy2006bec, sheehy2015fulde}, but first order transitions to the normal state are seen in the strongest pairing regime $Ut=0.1$ within numerical accuracy. An overall trend is that the critical Zeeman fields $h_{\textrm{c1}}$ for the USF-FMC transition and $h_{\textrm{c2}}$ for the final transition to the normal state both increase with the pairing interaction strength $U^{-1}$, but their difference $h_{\textrm{c2}} - h_{\textrm{c2}}$ slightly decreases -- indicating that a certain intermediate coupling is most optimal for the FMC states.

A notable trend among the PDW states is that ``diagonal'' orders ${\bf Q}=2\pi(f,f)$ appear at sufficiently large densities. The lowest chemical potential of such ``diagonal'' orders is pushed to higher $\mu$ values toward half filling ($\mu=4t$) by weakening the pairing interaction (increasing $U$). Therefore, stronger interactions favor and enhance the ``diagonal'' PDW states near half filling. In most cases, the PWS bordering the normal state at highest Zeeman fields is commensurate within numerical accuracy, living at ${\bf Q}=(\pi,0)$. At this wavevector, there is no distinction between the PWS and PDW states, and the condensate by itself does not violate the time-reversal symmetry.

\begin{figure}[!]
\subfigure[{}]{\raisebox{0.2in}{\includegraphics[height=1.55in]{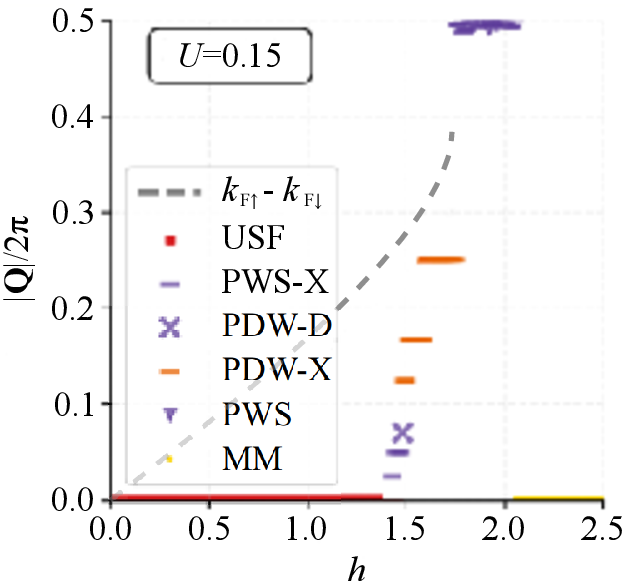}}}
\subfigure[{}]{\raisebox{0.2in}{\includegraphics[height=1.55in]{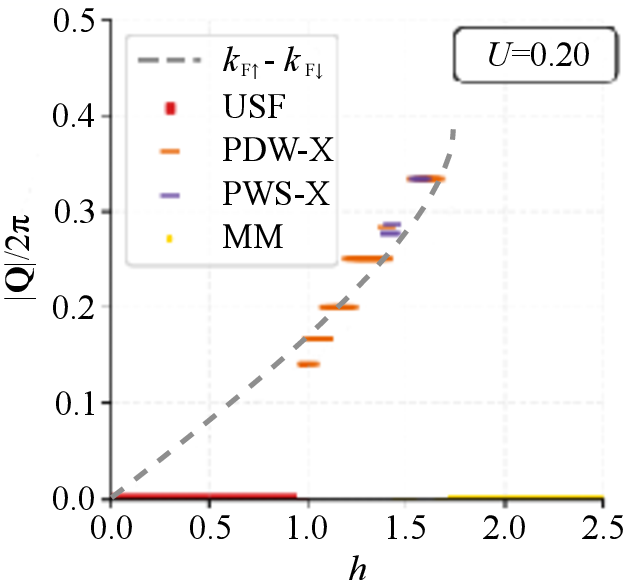}}}
\caption{\label{fig:compare-Q-evolution}Evolution of the ordering wavevector ${\bf Q}$ with the Zeeman field at $\mu=1.75$. (a) Strong coupling does not care about the Fermi surface even at low particle density; (b) Weaker coupling -- $|{\bf Q}|$ is in agreement with expetations from the Fermi surface gometry. $\mu,h,U$ are given in the units $t=a=1$.}
\end{figure}

The magnitude $|{\bf Q}|$ of the ordering wavevector is a good indicator of the interaction strength regime. In the weak coupling limit, one expects the FFLO states to order at wavevectors dictated by the shape of the Fermi surface. The difference of Fermi wavevectors along the direction ${\bf Q}=(q_x,0)$ is naively estimated from (\ref{Dispersion}) as
\begin{equation}
k_{\textrm{F}\uparrow}-k_{\textrm{F}\downarrow} = \arccos\left(1-\frac{\mu+h}{2t}\right)-\arccos\left(1-\frac{\mu-h}{2t}\right) \ ,
\end{equation}
and Figure \ref{fig:compare-Q-evolution} compares it with the actual ordering wavevectors of the numerically obtained states. The interaction coupling $Ut=0.2$ follows the expectations from the weak-coupling regime, but $Ut\le0.15$ is clearly in the strong-coupling limit. 

\begin{figure}
\includegraphics[height=2.5in]{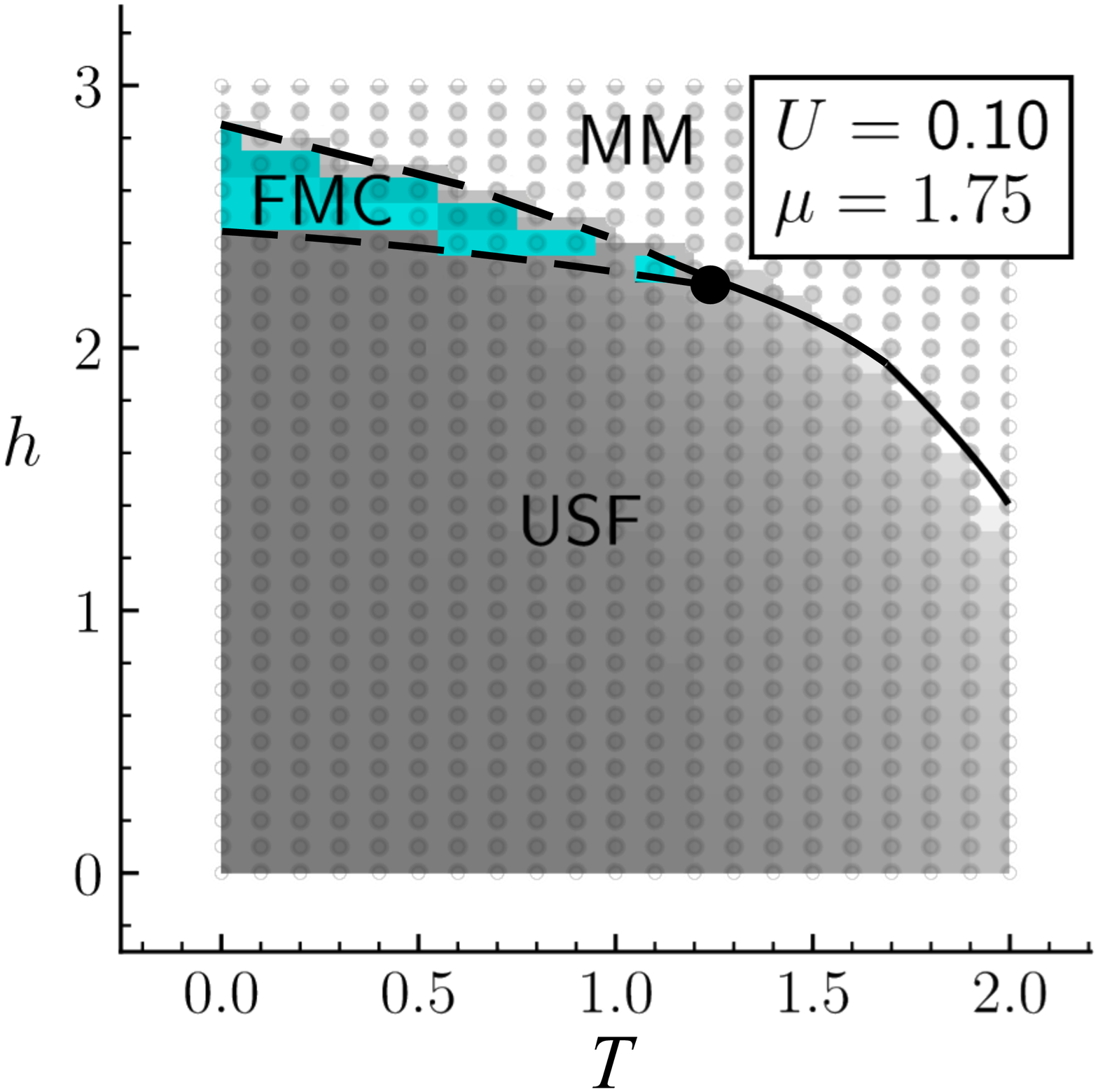} \hskip 0.25in
\includegraphics[height=2.5in]{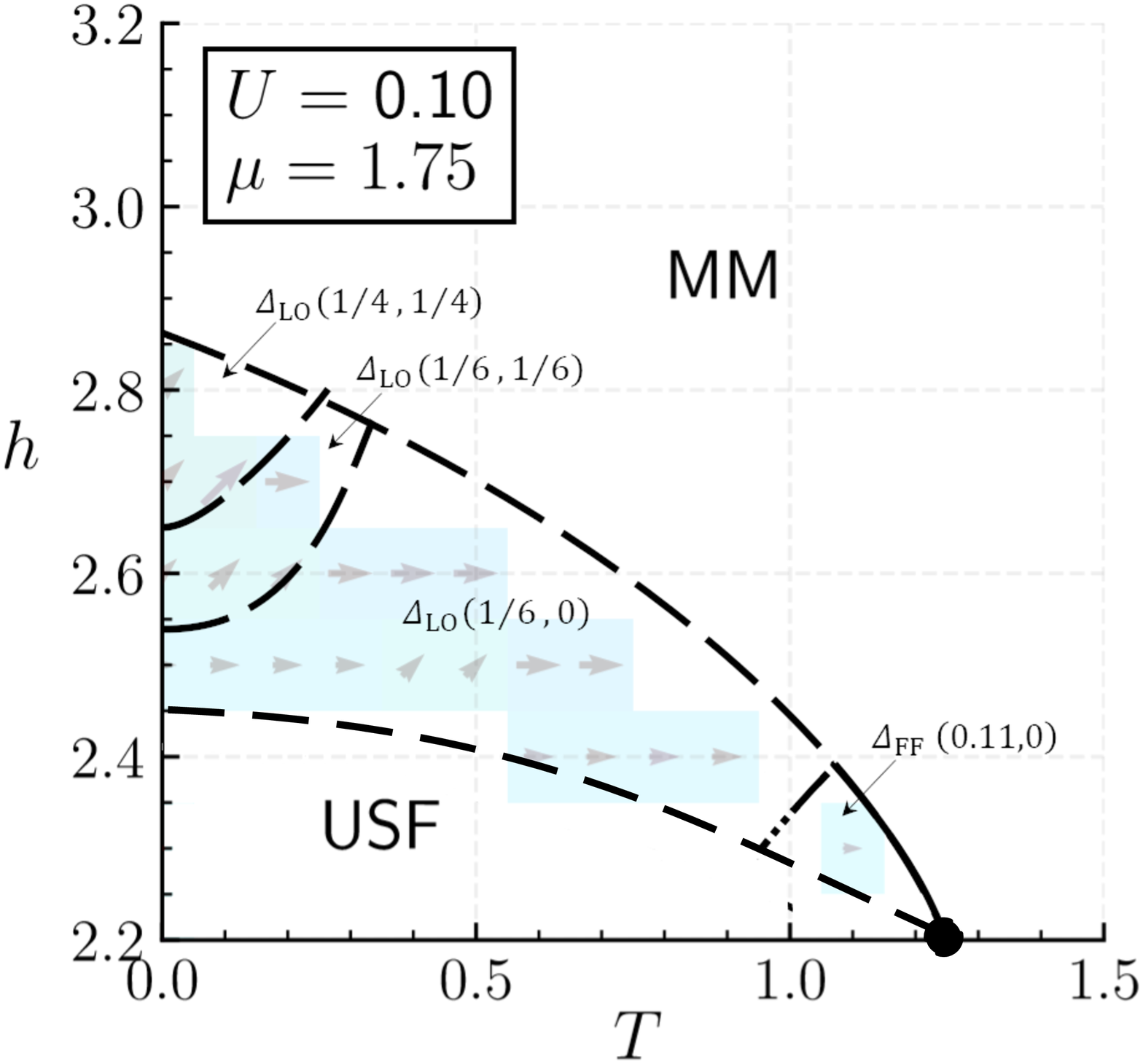}
\caption{\label{fig:phdiag-NonZero-010}The mean-field phase diagram at finite temperatures, plotted with the same conventions as Fig.\ref{fig:phdiag:mu-h} and Fig.\ref{fig:phdiag:mu-h:fmc}. The FMC-MM and USF-FMC transition lines meet at the marked Lifshitz critical point \cite{Koponen2007}. The parameters $\mu,h,T,U$ are given in the units $t=a=1$.}
\end{figure}

Fig.\ref{fig:phdiag-NonZero-010} is the mean-field phase diagram of FMC at finite temperatures. The continuous U(1) symmetry of the superfluid states is restored due to fluctuations (with surviving algebraic correlations below the Kosterlitz-Thouless transition), but the discrete lattice symmetries remain spontaneously broken in the PDW states. The familiar physical picture that emerges is similar to the zero temperature case where pairing at zero wavevector is able to withstand small $h$ until a region of FMC provides a compromise between the Zeeman field and the interaction coupling. Proliferating thermal fluctuations are found to continuously transform the incommensurate PWS state at highest Zeeman fields toward smaller condensation wavevectors $|{\bf Q}|$. The ordering wavevector of the PDW states is rigidly locked to the lattice and not flexible; less stable PDW condensates with smaller $|{\bf Q}|$ are seen to give way to the USF as temperature rises. Only the uniform superfluid USF is stable above a certain critical temperature.

\subsection{Fluctuation effects beyond the mean-field approximation}\label{secFluct}

\begin{figure}
\includegraphics[height=1.6in]{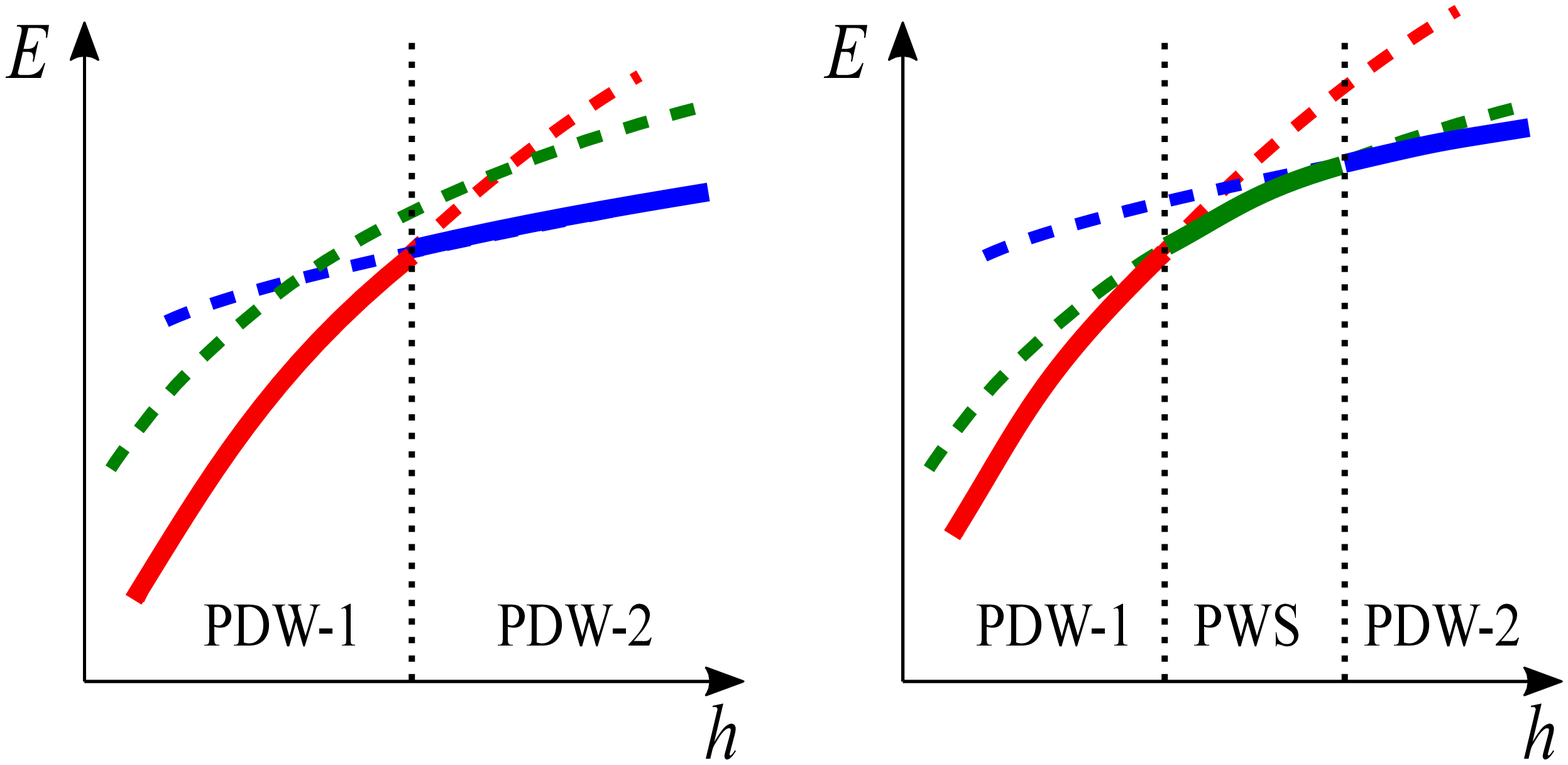}
\caption{\label{fig:incommPWS}The scenario for the appearance of incommensurate PWS between competing PDW states. The diagrams illustrate a typical evolution of energy with the Zeeman field for three competing states, two PDW states (red and blue) and another PWS state (green). The lowest energy state is the ground state, and the left diagram is the mean-field prediction. The presence of quantum fluctuations beyond the mean-field approximation is expected to raise the energy of all ordered states, but more so of the states that break the translational symmetry (since their defining positional order will be blurred out and weakened) -- illustrated in the right diagram. This makes it possible for a PWS to win in the intermediate region of the phase diagram between the two adjacent PDW orders.}
\end{figure}

The mean-field approximation neglects all quantum fluctuations of the order parameter. Such fluctuations generally shrink the condensate regions in the phase diagram, and possibly eliminate some phases found at the mean-field level in the given microscopic model. Another notable possibility is the appearance and spread of incommensurate PWS. Our finite temperature results hint at the ability of fluctuations to stabilize incommensurate PWS near the transition to the normal state. Furthermore, we expect that the translation-symmetry-breaking PDW states are more sensitive to quantum fluctuations than the translation-invariant PWS states -- the PDW energy should increase more due to fluctuations when their defining positional order washes out. Consequently, incommensurate PWS may show up as intermediate phases between various adjacent PDW phases of the mean-field phase diagram (see Fig.\ref{fig:incommPWS}).

All mean-field phases are still stable at zero temperature in a more general set of models with the same symmetries and degrees of freedom. However, the model (\ref{modelHamiltonian}) arises from the mean-field treatment of the attractive Hubbard model, whose special properties at half filling invalidate the present mean-field approach. Including quantum fluctuations in the analysis amounts to allowing $\Delta_{\bf q}$ in (\ref{modelHamiltonian}) to fluctuate. $\Delta$ is actually a Hubbard-Stratonovich field in the path-integral, and integrating it out yields the attractive Hubbard model (AHM). This model can be written in real space as
\begin{eqnarray}\label{AHM}
H_{\textrm{AHM}} &=& -t\sum_{\langle ij\rangle}(c_{i\sigma}^{\dagger}c_{j\sigma}^{\phantom{\dagger}}+c_{j\sigma}^{\dagger}c_{i\sigma}^{\phantom{\dagger}})
-\mu\sum_{i}n_{i}^{\phantom{x}}-h\sum_{i}S_{i}^{z} \nonumber \\
&& -u\sum_{i}\left(n_{i\uparrow}^{\phantom{x}}-\frac{1}{2}\right)\left(n_{i\downarrow}^{\phantom{x}}-\frac{1}{2}\right)
\end{eqnarray}
in the large $u=U^{-1}$ limit at half filling ($\mu=h=0$ and $\sum_i n_i = N$, where $n_i^{\phantom{\dagger}} = n_{i\uparrow}^{\phantom{\dagger}}+n_{i\downarrow}^{\phantom{\dagger}}$, $S_i^z = \frac{1}{2}(n_{i\uparrow}^{\phantom{\dagger}}-n_{i\downarrow}^{\phantom{\dagger}})$, $n_{i\sigma}^{\phantom{\dagger}} = c_{i\sigma}^\dagger c_{i\sigma}^{\phantom{\dagger}}$ and $N$ is the number of sites on the lattice). The mapping
\begin{eqnarray}
c_{i\uparrow}^{\phantom{\dagger}}\to f_{i\uparrow}^{\phantom{\dagger}}\quad,\quad c_{i\downarrow}^{\dagger}\to(-1)^{i}f_{i\downarrow}^{\phantom{\dagger}} \ ,
\end{eqnarray}
where $(-1)^i$ is $+1$ on the A-sublattice and $-1$ on the B-sublattice of the bipartite square lattice, transforms the AHM to the equivalent repulsive Hubbard model (RHM) with a modified chemical potential $\mu'$ and Zeeman field $h'$ given by:
\begin{equation}
\mu'=\frac{h}{2}\quad,\quad\mu=\frac{h'}{2} \ .
\end{equation}
The $f$-fermions of the RHM are localized at half filling in the large $u$ limit, so the ensuing low-energy dynamics is captured by a Heisenberg model whose SU(2) symmetry is broken in the Neel antiferromagnetic ground state. Since the RHM spin maps to the AHM density, the ground state generally carries a checkerboard density wave of the original $c$-fermions. This is combined with superfluidity of $c$-fermions, given that the broken symmetry is SU(2) and the density wave breaks only a discrete lattice symmetry. The ground state of the AHM at half filling has an SU(2) degeneracy spanning density wave and superfluid orders. The mean-field ``diagonal'' PDW states at large wavevectors near half filling can be viewed as precursors of this state, but the mean-field approximation becomes less reliable.

The mentioned SU(2) degeneracy is fragile, so the ground state of AHM can be tipped over to a pure superfluid or a Mott insulator of localized particles by weak perturbations. The chemical potential $\mu$ that takes AHM away from half filling is actually a perturbation that stabilizes a pure superfluid. In the language of RHM, $\mu=h'/2$ corresponds to a Zeeman field which first makes the Neel order rotate into the $xy$-plane (perpendicular to the external field), letting it spontaneously break the remaining U(1) in-plane rotation symmetry, then somewhat tilts the spins toward the $z$-axis to build magnetization. Since the obtained magnetization is uniform, the mapped particle density of the AHM is uniform as well, but the superfluidity remains.

A different perturbation will facilitate a spontaneous breaking of the original SU(2) symmetry into a checkerboard Mott insulator of localized particles:
\begin{equation}\label{AHMrep}
H'_{\textrm{AHM}}=V\sum_{\langle ij\rangle}n_{i}^{\phantom{\dagger}}n_{j}^{\phantom{\dagger}} \quad\leftrightarrow\quad H'_{\textrm{RHM}}=4V\sum_{\langle ij\rangle}\mathcal{S}_{i}^{z}\mathcal{S}_{j}^{z} \ .
\end{equation}
This is a repulsion between two particles on nearest-neighbor bonds in the AHM model, equivalent to the $z$-axis Ising coupling in the RHM (at half filling). In the RHM language, all spins of the already established Neel state will simply rotate to become parallel or antiparallel to the $z$-axis. The equivalent state of $c$-fermions in the AHM is a pure density wave without superfluidity: all particles are localized. Given the large interaction $u$, this state is actually a checkerboard Mott insulator of tightly bound Cooper pairs.

Perturbations such as (\ref{AHMrep}) are natural in realistic systems since particle interactions always have a finite range, and the short-range interactions between tight Cooper pairs in the strong coupling BEC regime are necessarily repulsive due to the Pauli exclusion that governs their fermionic constituents. Therefore, phases with localized Cooper pairs are a possible outcome of quantum fluctuations near half filling in the strong coupling limit. The physical picture of fluctuations considered so far has neglected the Zeeman field. A sufficiently large Zeeman field $h$ (in the AHM language) would eventually destabilize a localized state by breaking up some Cooper pairs and building up a spin-polarized background of unpaired fermions. It is possible for the particles of the ensuing state to remain localized, giving rise to unconventional polarized Mott insulators that break the lattice symmetries at a variety of ordering wavevectors. On the other hand, the reduced density of Cooper pairs will make them more vulnerable to fluctuations and susceptible to FFLO condensation.

\section{Conclusions}\label{secConclusions}

We studied the mean-field phase diagram of a simple two-dimensional model of attractively interacting fermions on the square lattice in the presence of a Zeeman field.  We discovered that the lattice stabilizes finite momentum condensates in a wide range of parameters. If the ground state is a uniform superfluid at zero Zeeman field, it can evolve through a sequence of first-order phase transitions separating commensurate pair density wave condensates as the Zeeman field is raised beyond a critical value. The superfluid state surviving at the highest Zeeman fields is typically a plane wave condensate, which becomes incommensurate at finite temperatures and terminates at the Lifshitz critical point. The final transition from such a state to the normal state (magnetized Fermi liquid) is found to be continuous. Strong interactions enhance the richness of finite momentum condensates and aid their stability within the mean-field approximation. However, fluctuation effects are important in this regime as well, and we speculate (based on indications from our work) that Cooper pair localization may take place near half filling.

\section{Acknowledgments}\label{secAck}

We are grateful for insightfull discussions with Ahmet Keles. This work was supported by the National Science Foundation grant PHY-1205571.

\appendix

\section{Notes on the numerical calculation}

Mean-field condensates are determined by (free) energy minimization under the variations of the superfluid order parameter. We consider order parameters in all three categories given in Section \ref{secPD}. In categories 1 and 2, we systematically examine the commensurate ordering wavevectors with fractional values $Q=2\pi f$, starting with $f=1/2$ and ending with $f=1/10$. In category 3, the ordering wavevector is varied continuously together with the order parameter. Variations of the order parameter (and the incommensurate ${\bf Q}$ in category 3) are generated with the simplex method of multi-variable function optimization. We minimize the free energy (\ref{FE}) in each category and pick the order parameter with the lowest free energy found.

\begin{figure}[!]
\subfigure[{}]{\raisebox{0.2in}{\includegraphics[height=1.9in]{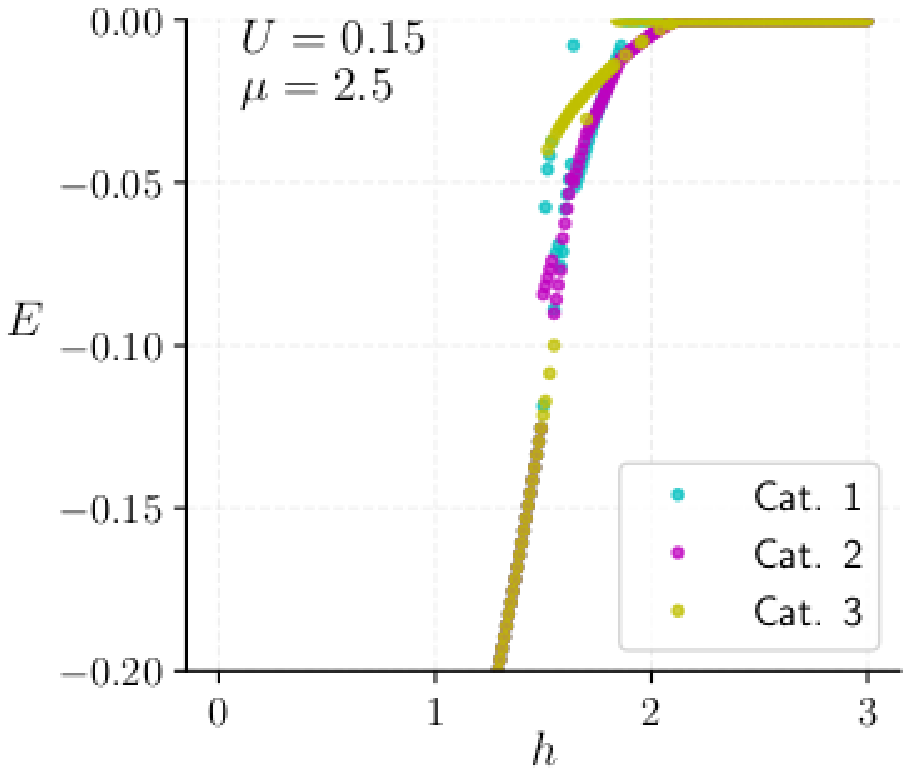}}}
\subfigure[{}]{\raisebox{0.2in}{\includegraphics[height=1.9in]{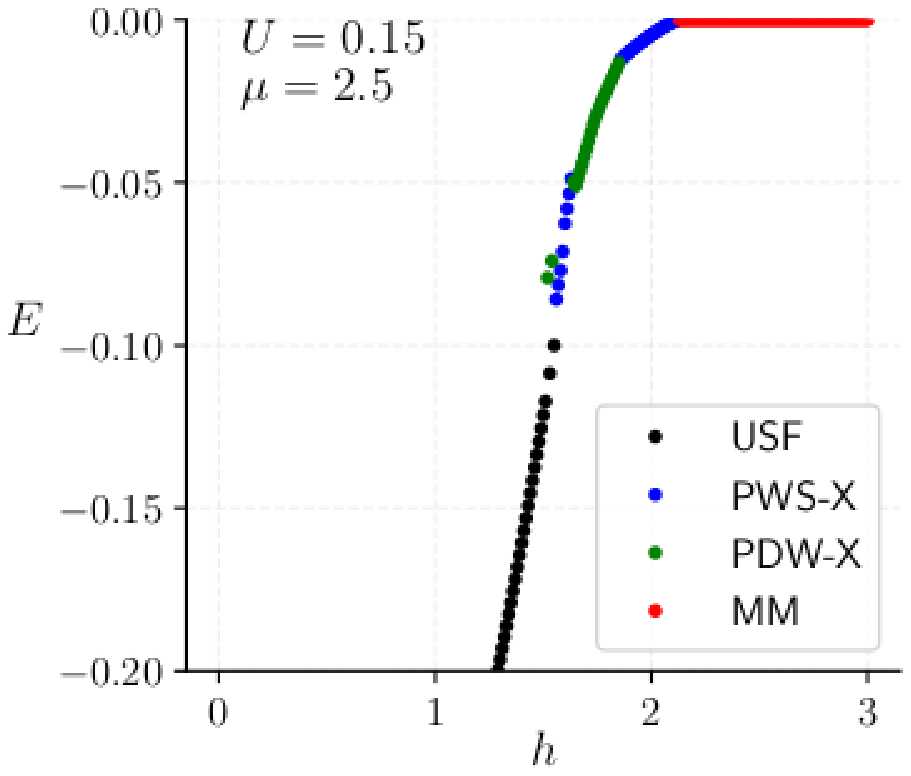}}}
\caption{\label{fig:metastable}A comparison of different mean-field metastable condensates by energy. (a) Raw data-points obtained in multiple runs of all category calculations; (b) minimum energy states across all runs. $\mu,h,U$ are given in the units $t=a=1$.}
\end{figure}

The process of free energy minimization starts with a ``seed'' order parameter and discovers a local minimum. The energy landscape can be complicated by multiple local minimums that correspond to metastable states. Fundamentally, there is no guarantee that the procedure finds the global free energy minimum. Therefore, we performed multiple runs with different random seeds. Collecting information about all found metastable states and observing the trends provides a certain confidence in the final conclusions about the equilibrium phase diagram. Fig.\ref{fig:metastable} illustrates a typical set of stable and metastable states obtained in our numerical calculations. Since the model has well-defined cut-offs, our numerical integration by the Legendre method is the only source of errors. We estimated these errors by varying the order of the fitting Legendre polynomial, and verified that they are smaller than the energy differences between metastable states.

\clearpage


%

\end{document}